\documentclass{sig-alternate-05-2015}
\usepackage{lcsect}
\usepackage{csquotes}
\usepackage{amsmath}
\usepackage{amssymb}
\usepackage{epsfig}
\usepackage{epstopdf}
\usepackage{algorithm}
\usepackage{algpseudocode}
\usepackage{graphicx}
\usepackage{array}
\usepackage{float}
\usepackage{balance}
\usepackage{url}
\usepackage{tabularx}
\usepackage[caption=false]{subfig}
\usepackage{blindtext}
\usepackage{etoolbox}
\makeatletter
\patchcmd{\maketitle}{\@copyrightspace}{}{}{}
\makeatother

\begin{document}


\doi{http://dx.doi.org/xx.xxxx/xxxxxxx.xxxxxxx}

\isbn{978-1-4503-3739-7/16/04}



%
\title{Using Hadoop for Large Scale Analysis on Twitter: A Technical Report}

\numberofauthors{3}
\author{
\alignauthor
Nikolaos Nodarakis \\
  \affaddr{Computer Engineering and Informatics Department, University of Patras}\\
  \affaddr{26500 Patras, Greece}\\
  \email{nodarakis@ceid.upatras.gr}
  \alignauthor
Spyros Sioutas\\
  \affaddr{Department of Informatics, Ionian University}\\
  \affaddr{49100 Corfu, Greece}\\
  \email{sioutas@ionio.gr}
  \and
  \alignauthor
Athanasios Tsakalidis\\
  \affaddr{Computer Engineering and Informatics Department, University of Patras}\\
  \affaddr{26500 Patras, Greece}\\
  \email{tsak@ceid.upatras.gr}
  \alignauthor
Giannis Tzimas \\
  \affaddr{Computer \& Informatics Engineering Department}\\	
  \affaddr{Technological Educational}
  \affaddr{Institute of Western Greece}
  \affaddr{26334 Patras, Greece}
  \email{tzimas@cti.gr}
}

\maketitle
\begin{abstract}
Sentiment analysis (or opinion mining) on Twitter data has attracted much attention recently. One of the system's key features, is the immediacy in communication with other users in an easy, user-friendly and fast way. Consequently, people tend to express their feelings freely, which makes Twitter an ideal source for accumulating a vast amount of opinions towards a wide diversity of topics. This amount of information offers huge potential and can be harnessed to receive the sentiment tendency towards these topics. However, since none can invest an infinite amount of time to read through these tweets, an automated decision making approach is necessary. Nevertheless, most existing solutions are limited in centralized environments only. Thus, they can only process at most a few thousand tweets. Such a sample, is not representative to define the sentiment polarity towards a topic due to the massive number of tweets published daily. In this paper, we go one step further and develop a novel method for sentiment learning in the MapReduce framework. Our algorithm exploits the hashtags and emoticons inside a tweet, as sentiment labels, and proceeds to a classification procedure of diverse sentiment types in a parallel and distributed manner. Moreover, we utilize Bloom filters to compact the storage size of intermediate data and boost the performance of our algorithm. Through an extensive experimental evaluation, we prove that our solution is efficient, robust and scalable and confirm the quality of our sentiment identification.
\end{abstract}

%
%


%
%

%
%




%
\section{Introduction}
\label{intro}
Twitter is one of the most popular social network websites and launched in 2006. Since then, it has grown at a very fast pace and at the time speaking numbers 316 million monthly active users, while 500 millions tweets are sent on a daily basis\footnote{\url{https://about.twitter.com/company} (Visited 19/9/2015)}. Naturally, it is a wide spreading instant messaging platform and people use it to get informed about world news, videos that have become viral, discussions over recently released products or technological advancements, etc. Inevitably, a cluster of different opinions, that carry rich sentiment information and concern a variety of entities or topics, is formed. Sentiment is defined as "A thought, view, or attitude, especially one based mainly on emotion instead of reason"\footnote{\url{http://www.thefreedictionary.com/sentiment}} and describes someone's  mood or judge towards a specific entity. User-generated content that captures sentiment information has proved to be valuable and its use is widespread among many internet applications and information systems, such as search engines.

Knowing the overall sentiment inclination towards a topic, provides very useful information and can be captivating in certain cases. For instance, Google would like to know what their users think about the latest Android 5.0 update, in order to proceed to further development and bug fixing until the operating system works smoothly and meets the needs of the users. Thus, it is clear that a concise sentiment analysis towards the topic during a time period is needed. Two of the most known websites that perform sentiment analysis on Twitter are Topsy\footnote{\url{http://topsy.com/}} and Sentiment140\footnote{\url{http://www.sentiment140.com/}}.

In the context of this work, we utilize \textit{hashtags} and \textit{emoticons} as sentiment labels to perform classification of diverse sentiment types. Hashtags are a convention for adding additional context and metadata to tweets. They are created by users as a way to categorize their message and/or highlight a topic and are extensively utilized in tweets \cite{Wang:2011:TSA:2063576.2063726}. Moreover, they provide the ability to people to search tweets that refer to a common subject. The creation of a hashtag is achieved by prefixing a word with a hash symbol (e.g. \#love). Emoticon refers to a digital icon or a sequence of keyboard symbols that serves to represent a facial expression, as {\tt :-)} for a smiling face\footnote{\url{http://dictionary.reference.com/browse/emoticon}}. Both, hashtags and emoticons, provide a fine-grained sentiment learning at tweet level which makes them suitable to be leveraged for opinion mining.

Although the problem of sentiment analysis has been studied extensively during recent years, existing solutions suffer from certain limitations. One problem is that the majority of approaches is bounded in centralized environments. Moreover, sentiment analysis is based on, it terms of methodology, natural language processing techniques and machine learning approaches. However, this kind of techniques are time-consuming and spare many computational resources. Consequently, at most a few thousand records can be processed by such techniques without exceeding the capabilities of a single server. Since millions of tweets are published daily on Twitter, it is more than clear that underline solutions are not sufficient. Consequently, high scalable implementations are required in order to acquire a much better overview of sentiment tendency towards a topic. Cloud computing technologies provide tools and infrastructure to create such solutions and manage the input data in a distributed way among multiple servers. The most popular and notably efficient tool is the MapReduce \cite{dean:ghem} programming model, developed by Google, for processing large-scale data.

In this paper, we propose MR-SAT: a novel \underline{M}ap\underline{R}educe Algorithm for Big Data \underline{S}entiment \underline{A}nalysis on \underline{T}witter implemented in Hadoop \cite{apache:hadoop, reilly}, the open source MapReduce implementation. Our algorithm exploits the hashtags and emoticons inside a tweet, as sentiment labels, in order to avoid the time-intensive manual annotation task. After that, we build the feature vectors of training and test set and proceed to a classification procedure in a fully distributed manner. Additionally, we encode features using Bloom filters to compress the storage space of the feature vectors. We adapt an existing MapReduce classification algorithm based on A$k$NN queries to achieve the desirable outcome. Through an extensive experimental evaluation we study various parameters that can affect the total computation cost and classification performance. We prove that our solution is efficient, robust and scalable and confirm the quality of our sentiment identification.

The rest of the paper is organized as follows: in Section 2 we discuss related work and in Section 3 we present how our algorithm works. More specifically, we explain how to build the feature vectors (for both the training and test dataset), we briefly describe the Bloom filter integration and display our A$k$NN based classification algorithm. After that, we proceed to the experimental evaluation of our approach in Section 4, while in Section 5 we conclude the paper and present future steps.

\section{Related Work}
\label{related_work}

The domain of sentiment analysis, or opinion mining, has been studied extensively in literature during decent years. Early studies focus on document level sentiment analysis concerning movie or product reviews \cite{Hu:2004:MSC:1014052.1014073, Zhuang:2006:MRM:1183614.1183625} and posts published on webpages or blogs \cite{Zhang:2007:ORB:1321440.1321555}. Respectively, many efforts have been made towards the sentence level sentiment analysis \cite{Wilson:2005:RCP:1220575.1220619, Wilson:2009:RCP:1618327.1618330, Yu:2003:TAO:1119355.1119372} which examines phrases and assigns to each one of them a sentiment polarity (positive, negative, neutral). A less investigated area is the topic-based sentiment analysis \cite{Lin:2009:JSM:1645953.1646003, Mei:2007:TSM:1242572.1242596} due to the difficulty to provide an adequate definition of topic and how to incorporate the sentiment factor into the opinion mining task.  

Many researchers confront the problem of sentiment analysis by applying machine learning approaches and/or natural language processing techniques. In \cite{Pang:2002:TUS:1118693.1118704}, the authors employ three machine learning techniques to classify movie reviews as positive or negative. On the other hand, Nasukawa and Yi \cite{Nasukawa:2003:SAC:945645.945658} investigate the proper identification of semantic relationships between the sentiment expressions and the subject, in order to enhance the accuracy of sentiment analysis within webpages and online articles. Their approach utilizes a syntactic parser and a sentiment lexicon. Moreover, Ding and Liu \cite{Ding:2007:ULR:1277741.1277921} propose a set of linguistic rules together with a new opinion aggregation function to detect sentiment orientations in online product reviews.

Nowadays, Twitter has received much attention for sentiment analysis, as it provides a source of massive user-generated content that captures a wide aspect of published opinions. In \cite{Barbosa:2010:RSD:1944566.1944571}, the authors propose a 2-step classifier that separates messages as subjective and objective, and further distinguishes the subjective tweets as positive or negative. Davidov et al. \cite{Davidov:2010:ESL:1944566.1944594} exploit the hashtags and smileys in tweets and evaluate the contribution of different features (e.g. unigrams) together with a $k$NN classifier. In this paper, we adopt this approach and create a parallel and distributed version of the algorithm for large scale Twitter data. Agarwal et al. \cite{Agarwal:2011:SAT:2021109.2021114} explore the use of a tree kernel model for detecting sentiment orientation in tweets. A three-step classifier is proposed in \cite{Jiang:2011:TTS:2002472.2002492} that follows a target-dependent sentiment classification strategy by incorporating target-dependent features and taking related tweets into consideration. Moreover, the authors in \cite{Wang:2011:TSA:2063576.2063726} perform a topic sentiment analysis in Twitter data through a graph-based model. A more recent approach \cite{Yamamoto:2014:REM:2684200.2684283}, investigates the role of emoticons for multidimensional sentiment analysis of Twitter by constructing a sentiment and emoticon lexicon. A large scale solution is presented in \cite{Khuc:2012:TBL:2245276.2245364} where the authors build a sentiment lexicon and classify tweets using a MapReduce algorithm and a distributed database model. Although the classification performance is quite good, the construction of sentiment lexicon needs a lot of time. Our approach is much simpler and, to our best knowledge, we are the first to present a robust large scale approach for opinion mining on Twitter data without the need of building a sentiment lexicon or proceeding to any manual data annotation.

\section{MR-SAT Approach}

We begin this section by providing a formal definition of the problem we try to tackle and then we present the features we use for sentiment classification. Finally, we describe our algorithm using pseudo-codes and proceed to a step by step explanation of each pseudo-code. Assume a set of hashtags $H = \{h_1, h_2, \dots , h_n\}$ and a set of emoticons $E = \{em_1, em_2, \dots , em_m\}$ associated with a set of tweets $T = \{t_1, t_2, \dots , t_l\}$ (training set). Each $t \in T$ carries only one sentiment label from $L = H \cup E$. This means that tweets containing more that one labels from $L$ are not candidates for $T$, since their sentiment tendency may be vague. Given a set of unlabelled tweets $TT = \{tt_1, tt_2, \dots , tt_k\}$ (test set), we aim to infer the sentiment polarities $p = \{p_1, p_2, \dots , p_k\}$ for $TT$, where $p_i \in L \cup \{neu\}$ and $neu$ means that the tweet carries no sentiment information. We build a tweet-level classifier $C$ and adopt a $k$NN strategy to decide the sentiment tendency $\forall tt \in TT$. We implement $C$ by adapting an existing MapReduce classification algorithm based on A$k$NN queries \cite{DBLP:conf/dexa/NodarakisPSTTT14}, as described in Subsection 3.3.

\subsection{Feature Description}

In this subsection, we present in details the features used in order to build classifier $C$. For each tweet we combine its features in one feature vector. We apply the features proposed in \cite{Davidov:2010:ESL:1944566.1944594} with some necessary modifications to avoid the production of an exceeding amount of calculations, thus boosting the running performance of our algorithm.

\subsubsection{Word and N-Gram Features}

We treat each word in a tweet as a binary feature. Respectively, we consider 2-5 consecutive words in a sentence as a binary n-gram feature. If $f$ is a word or n-gram feature, then $w_f = \frac{N_f}{count(f)}$ is the weight of $f$ in the feature vector, $N_f$ is the number of times $f$ appear in the tweet and $count(f)$ declares the count of $f$ in the Twitter corpus. Consequently, rare words and n-grams  have a higher weight than common words and have a greater effect on the classification task. Moreover, we consider sequences of two or more punctuation symbols as word features. Unlike what authors propose in \cite{Davidov:2010:ESL:1944566.1944594}, we do not include the substituted meta-words for URLs, references and hashtags (URL, REF and TAG respectively) as word features (see and Section 4). Also, the common word RT, which means "retweet", does not constitute a feature. The reason for omission of these words from the feature list lies in the fact that they appear in the majority of tweets inside the dataset. So, their contribution as features is negligible, whilst they lead to a great computation burden during the classification task.

\subsubsection{Pattern Features}

This is the main feature type and we apply the pattern definitions given in \cite{Davidov:2006:EUD:1220175.1220213} for automated pattern extractions. We classify words into three categories: high-frequency words (HFWs), content words (CWs) and regular words (RWs). A word whose corpus frequency is more (less) than $F_H$ ($F_C$) is considered to be a HFW (CW). The rest of the words are characterized as RWs. The word frequency is estimated from the training set rather than from an external corpus. In addition, we treat as HFWs all consecutive sequences of punctuation characters as well as URL, REF, TAG and RT meta-words for pattern extraction. We define a pattern as an ordered sequence of HFWs and slots for content words. The upper bound for $F_C$ is set to 1000 words per million and the lower bound for $F_H$ is set to 100 words per million. Observe that the $F_H$ and $F_C$ bounds allow overlap between some HFWs and CWs. To address this issue, we follow a simple strategy as described next. Assume $fr$ is the frequency of a word in the corpus; if $fr \in \left(F_H, \frac{F_H + F_C}{2}\right)$ the word is classified as HFW, else if $fr \in \left[\frac{F_H + F_C}{2}, F_C\right)$ the word is classified as CW.

We seek for patterns containing 2-6 HFWs and 1-5 slots for CWs. Moreover, we require patterns to start and to end with a HFW, thus a minimal pattern is of the form [HFW][CW slot][HFW]. Additionally, we allow approximate pattern matching in order to enhance the classification performance. Approximate pattern matching is the same as exact matching, with the difference that an arbitrary number of RWs can be inserted between the pattern components. Since the patterns can be quite long and diverse, exact matches are not expected in a regular base. So, we permit approximate matching in order to avoid large sparse feature vectors. The weight $w_p$ of a pattern feature $p$ is defined as $w_p = \frac{N_p}{count(p)}$ in case of exact pattern matching and as $w_p = \frac{\alpha \cdot N_p}{count(p)}$ in case of approximate pattern matching, where $\alpha = 0.1$ in all experiments.

\subsubsection{Punctuation Features}

The last feature type is divided into five generic features as follows: 1) tweet length in words, 2) number of exclamation mark characters in the tweet, 3) number of question mark characters in the tweet, 4) number of quotes in the tweet and 5) number of capital/capitalized words in the tweet. The weight $w_p$ of a punctuation feature $p$ is defined as $w_p = \frac{N_{p}}{M_{p} \cdot \left( M_w + M_{ng} + M_{pa}\right) / 3}$, where $N_{p}$ is the number of times feature $p$ appears in the tweet, $M_{p}$ is the maximal observed value of $p$ in the twitter corpus and $M_w, M_{ng}, M_{pa}$ declare the maximal values for word, n-gram and pattern feature groups, respectively. So, $w_p$ is normalized by averaging the maximal weights of the other feature types.

\subsection{Bloom Filter Integration}

Bloom filters are data structures proposed by Bloom \cite{Bloom:1970:STH:362686.362692} for checking element membership in any given set. A Bloom filter is a bit vector of length $z$, where initially all the bits are set to 0. We can map an element into the domain between 0 and $z-1$ of the Bloom filter, using $q$ independent hash functions $hf_1, hf_2, ..., hf_q$. In order to store each element $e$ into the Bloom filter, $e$ is encoded using the $q$ hash functions and all bits having index positions $hf_j(e)$ for $1 \leq j \leq q$ are set to 1. 

Bloom filters are quite useful and they compress the storage space needed for the elements, as we can insert multiple objects inside a single Bloom filter. In the context of this work, we employ Bloom filters to transform our features to numbers, thus reducing the space needed to store our feature vectors.  More precisely, instead of storing a feature we store the index positions in the Bloom filter that are set to 1. Nevertheless, it is obvious that the usage of Bloom filters may impose errors when checking for element membership, since two different elements may end up having exactly the same bits set to 1. The error probability is decreased as the number of bits and hash functions used grows. As shown in the experimental evaluation, the side effects of Bloom filters are negligible and boost the performance of our algorithm.

\subsection{kNN Classification Algorithm}

In order to assign a sentiment label for each tweet in $TT$, we apply a $k$NN strategy. Initially, we build the feature vectors for all tweets inside the training and test datasets ($F_T$ and $F_{TT}$ respectively). Then, for each feature vector $u$ in $F_{TT}$ we find all the feature vectors in $V \subseteq F_T$ that share at least one word/n-gram/pattern feature with $u$ (matching vectors). After that, we calculate the Euclidean distance $d(u,v), \forall v \in V$ and keep the $k$ lowest values, thus forming $V_k \subseteq V$ and each $v_i \in V_k$ has an assigned sentiment label $L_i, 1 \leq i \leq k$. Finally, we assign $u$ the label of the majority of vectors in $V_k$. If no matching vectors exist for $u$, we assign a "neutral" label. We build $C$ by adjusting an already implemented A$k$NN classifier in MapReduce to meet the needs of opinion mining problem.

\subsection{Algorithmic Description}

In this subsection, we describe in detail the sentiment classification process as implemented in the Hadoop framework. We adjust an already implemented MapReduce A$k$NN classifier to meet the needs of opinion mining problem. Our approach consists of a series of four MapReduce jobs, with each job providing input to the next one in the chain. These MapReduce jobs can be summarized as follows: 1) \textbf{Feature Extraction:} Extract the features from all tweets in $T$ and $TT$, 2) \textbf{Feature Vector Construction:} Build the feature vectors $F_T$ and $F_{TT}$ respectively, 3) \textbf{Distance Computation:} For each vector $u \in F_{TT}$ find the matching vectors (if any exist) in $F_T$, calculate the Euclidean distance $d(u,v), \forall v \in V$ and form $V_k \subseteq V$, 4) \textbf{Sentiment Classification:} Assign a sentiment label $\forall tt \in TT$.

The records provided as input to our algorithm have the format $<$tweet\textunderscore{id}, class, text $>$, where \texttt{class} refers either to a sentiment label for tweets in $T$ either to a no-sentiment flag for tweets in $TT$. In the following subsections, we describe each MapReduce job separately and analyze the Map and Reduce functions that take place in each one of them. 

\subsubsection{Feature Extraction}

\begin{algorithm}
\floatname{algorithm}{MapReduce Job}
\newcommand{\algorithmname}{MapReduce Job}
\renewcommand{\thealgorithm}{1}
\caption{\null}
\begin{algorithmic}[1]
\Function{Map}{$k1, v1$}
\State $t\textunderscore{id} = \textbf{getId}(v1); class = \textbf{getClass}(v1);$
\State $features = \textbf{getFeatures}(v1);$
\ForAll {$f \in features$} $//$ BF is BloomFilter
\State $\textbf{output}(\textbf{BF}(f.text),<t\textunderscore{id}, f.count, class>);$
\EndFor
\EndFunction
\Statex
\Function{Reduce}{$k2, v2$}
\State $feature\textunderscore{freq} = 0;$
\ForAll {$v \in v2$}
\State $feature\textunderscore{freq} = feature\textunderscore{freq} + v.count;$
\EndFor
\State $l = List\{\};$
\ForAll {$v \in v2$}
\State $weight = v.count / feature\textunderscore{freq};$
\State $l.add(\textbf{new} Record(v.t\textunderscore{id}, weight, v.class));$
\EndFor
\State $\textbf{output}(k2, l);$
\EndFunction
\end{algorithmic}
\end{algorithm}

In this MapReduce job, we extract the features, as described in Subsection 3.1, of tweets in $T$ and $TT$ and calculate their weights. The output of the job is an inverted index, where the key is the feature itself and the value is a list of tweets that contain it. In the MapReduce Job 1 pseudo-code, we sum up the Map and Reduce functions of this process.

The \textit{Map} function takes as input the records from $T$ and $TT$, extracts the features of tweets. Afterwards, for each feature it outputs a key-value record, where the feature itself is the key and the value consists of the id of the tweet, the class of the tweet and the number of times the feature appears inside the sentence. The \textit{Reduce} function receives the key-value pairs from the Map function and calculates the weight of a feature in each sentence. Then, it forms a list $l$ with the format $<t_1, w_1, c_1 : ... : t_x, w_x, c_x>$, where $t_i$ is the id of the $i$-th tweet, $w_i$ is the weight of the feature for this tweet and $c_i$ is its class. For each key-value pair, the Reduce function outputs a record where the feature is the key and the value is list $l$.

\subsubsection{Feature Vector Construction} 

In this step, we build the feature vectors $F_T$ and $F_{TT}$ needed for the subsequent distance computation process. To achieve this, we combine all features of a tweet into one single vector. Moreover, $\forall tt \in TT$ we generate a list ($training$) of tweets in $T$ that share at least one word/n-gram/pattern feature. The Map and Reduce functions are outlined in the following MapReduce Job 2 pseudo-code.

\begin{algorithm}[h]
\floatname{algorithm}{MapReduce Job}
\newcommand{\algorithmname}{MapReduce Job}
\renewcommand{\thealgorithm}{2}
\caption{\null}
\begin{algorithmic}[1]
\Function{Map}{$k1, v1$}
\State $f= \textbf{getFeature}(v1); t\textunderscore{list} = \textbf{getTweetList}(v1);$
\State $test = training = List\{\};$
\ForAll {$t \in t\textunderscore{list}$}
\State $\textbf{output}(t.t\textunderscore{id}, <f, t.weight>);$
\If {$t.class \neq NULL$}
\State $training.add(\textbf{new} Record(t.t\textunderscore{id}, t.class));$
\Else 
\State $test.add(\textbf{new} Record(t.t\textunderscore{id}, t.class));$
\EndIf
\EndFor
\ForAll {$t \in test$}
\State $\textbf{output}(t.t\textunderscore{id}, training);$
\EndFor
\EndFunction
\Statex
\Function{Reduce}{$k2, v2$}
\State $features = training = List\{\};$
\ForAll {$v \in v2$}
\If {$v \textbf{ instanceOf } List$}
\State $training.addAll(v);$
\Else
\State $features.add(v);$
\EndIf
\EndFor
\If {$training.size() > 0$}
\State $\textbf{output}(k2, <training, features>);$
\Else
\State $\textbf{output}(k2, features);$
\EndIf
\EndFunction
\end{algorithmic}
\end{algorithm}

Initially, the \textit{Map} function separates $\forall f \in F$ the tweets that contain $f$ into two lists, $training$ and $test$ respectively. Also, $\forall f \in F$ it outputs a key-value record, where the key is the tweet id that contains $f$ and the value consists of $f$ and weight of $f$. Next, $\forall v \in test$ it generates a record where the key is the id of $v$ and the value is the $training$ list. The \textit{Reduce} function gathers key-value pairs with the same key and build $F_T$ and $F_{TT}$. For each tweet $t \in T$ ($tt \in TT$) it outputs a record where key is the id of $t$ ($tt$) and the value is its feature vector (feature vector together with the $training$ list).

\subsubsection{Distance Computation}

In MapReduce Job 3, we create pairs of matching vectors between $F_T$ and $F_{TT}$ and compute their Euclidean distance. The Map and Reduce functions are depicted in the pseudo-code that follows.

\begin{algorithm}[h]
\floatname{algorithm}{MapReduce Job}
\newcommand{\algorithmname}{MapReduce Job}
\renewcommand{\thealgorithm}{3}
\caption{\null}
\begin{algorithmic}[1]
\Function{Map}{$k1, v1$}
\State $t\textunderscore{ids} = \textbf{getTrainingIds}(v1); v = \textbf{getVector}(v1);$
\State $t\textunderscore{id} = \textbf{getId}(v1);$
\If {$t\textunderscore{ids}.size() > 0$}
\ForAll {$u \in t\textunderscore{ids}$}
\State $\textbf{output}(u.t\textunderscore{id}, <u.class,t\textunderscore{id}, v>);$
\EndFor
\Else
\State $\textbf{output}(t\textunderscore{id}, v);$
\EndIf
\EndFunction
\Statex
\Function{Reduce}{$k2, v2$}
\State $ttv = List\{\}; tv = NULL$
\ForAll {$v \in v2$}
\If {$v.class \neq NULL$}
\State $ttv.add(v);$
\Else
\State $tv = v;$
\EndIf 
\EndFor
\ForAll {$tt \in ttv$}
\State $\textbf{ouput}(tt.t\textunderscore{id}, <tv.t\textunderscore{id}, tv.class, d(tt, tv)>);$
\EndFor
\EndFunction
\end{algorithmic}
\end{algorithm}

For each feature vector $u \in F_{TT}$, the \textit{Map} function outputs all pairs of vectors $v$ in $training$ list of $u$. The output key-value record has as key the id of $v$ and the value consists of the class of $v$, the id of $u$ and the $u$ itself. Moreover, the Map function outputs all feature vectors in $F_T$. The \textit{Reduce} function concentrates $\forall v \in F_T$ all matching vectors in $F_{TT}$ and computes the Euclidean distances between pairs of vectors. The Reduce function produces key-value pairs where the key is the id of $u$ and the value comprises of the id of $v$, its class and the Euclidean distance $d(u, v)$ between the vectors.

\subsubsection{Sentiment Classification}

This is the final step of our proposed approach. In this job, we aggregate for all feature vectors $u$ in the test set, the $k$ vectors with the lowest Euclidean distance to $u$, thus forming $V_k$. Then, we assign to $u$ the label (class) $l \in L$ of the majority of $V_k$, or the $neu$ label if $V_k = \emptyset$. The MapReduce Job 4 pseudo-code is given below.

\begin{algorithm}[h]
\floatname{algorithm}{MapReduce Job}
\newcommand{\algorithmname}{MapReduce Job}
\renewcommand{\thealgorithm}{4}
\caption{\null}
\begin{algorithmic}[1]
\Function{Map}{$k1, v1$}
\State $t\textunderscore{id} = \textbf{getTweetId}(v1); val = \textbf{getValue}(v1);$
\State $\textbf{output}(t\textunderscore{id}, val);$
\EndFunction
\Statex
\Function{Reduce}{$k2, v2$}
\State $l\textunderscore{k} = \textbf{getKNN}(v2);$
\State $H = HashMap<Class, Occurences>\{\};$
\State $H = \textbf{findClassOccur}(l\textunderscore{k});$
\State $max = 0; maxClass = \textbf{null};$
\ForAll {$entry \in H$}
\If {$entry.occur > max$}
\State $max = entry.occur;$
\State $maxClass = entry.class;$
\EndIf
\EndFor
\State $\textbf{output}(k2, maxClass);$
\EndFunction
\end{algorithmic}
\end{algorithm}

The \textit{Map} function is very simple and it just dispatches the key-values pairs it receives to the \textit{Reduce} function. For each feature vector $u$ in the test set, the Reduce function keeps the $k$ feature vectors with the lowest distance to $v$ and then estimates the prevailing sentiment label $l$ (if exists) among these vectors. Finally, it assigns to $u$ the label $l$.

\section{Experimental Evaluation}

In this section, we conduct a series of experiments to evaluate the performance of our method under many different perspectives. More precisely, we take into consideration the effect of $k$ and Bloom filters, the space compaction ratio and the size of the dataset in the performance of our solution.

Our cluster includes 4 computing nodes (VMs), each one of which has four 2.4\,GHz CPU processors, 11.5\,GB of memory, 45\,GB hard disk and the nodes are connected by 1 gigabit Ethernet. On each node, we install Ubuntu 14.04 operating system, Java 1.7.0\textunderscore{51} with a 64-bit Server VM, and Hadoop 1.2.1. Moreover, we apply the following changes to the default Hadoop configurations: the replication factor is set to 1; the maximum number of Map and Reduce tasks in each node is set to 3 (consequently we set the number of Reduce tasks to 12), the DFS chunk size is 64\,MB and the size of virtual memory for each Map and Reduce task is set to 512\,MB.

We evaluate our method using two Twitter datasets (one for hashtags and one for emoticons) we have collected through the Twitter Search API\footnote{\url{https://dev.twitter.com/rest/public/search}} between November 2014 to August 2015. We have used two human judges to create a list of hashtags and a list emoticons that express strong sentiment (e.g \#bored and {\tt :)}). We performed some experimentation to exclude from the lists the hashtags and emoticons that either were abused by twitter users or returned a very small number of tweets. We ended up with a list of 13 hashtags and a list of 4 emoticons. We preprocessed the datasets we collected and kept only the English tweets which contained 5 or more proper English words\footnote{To identify the proper English word we used an available WN-based English dictionary} and do not contain two or more hashtags or emoticons from the aforementioned lists. Moreover, during preprocessing we have replaced URL links, hashtags and references by URL/REF/TAG meta-words as stated in \cite{Davidov:2010:ESL:1944566.1944594}. The final hashtags dataset contains 942188 tweets (72476 tweets for each class) and the final emoticons dataset contains 1337508 tweets (334377 tweets for each class). In both datasets, hashtags and emoticons are used as sentiment labels and for each sentiment label there is an equal amount of tweets. Finally, we produced two no-sentiment datasets by randomly sampling 72476 and 334377 tweets with no hashtags/emoticons from the dataset used in \cite{Cheng:2010:YYT:1871437.1871535} and is publicly available\footnote{\url{https://archive.org/details/twitter_cikm_2010}}. We assume that such random samples are unlikely to contain a significant amount of sentiment sentences. These datasets are used for the binary classification experiments (see Section 4.1).

We assess the classification performance of our algorithm using the 10-fold cross validation method and measuring the harmonic f-score. For the Bloom filter construction we use 999 bits and 3 hash functions. In order to avoid a significant amount of computations that greatly affect the running performance of the algorithm, we define a weight threshold $w = 0.005$ for feature inclusion in the feature vectors. In essence, we eliminate the most frequent words that have no substantial contribution to the final outcome.

\subsection{Classification Performance}

In this subsection we measure the classification performance of our solution using the harmonic f-score. We use two experimental settings, the multi-class classification and the binary classification settings. Under multi-class classification we attempt to assign a single label to each of vectors in the test set. In the binary classification experiments, we classified a sentence as either appropriate for a particular label or as not bearing any sentiment. As stated and in \cite{Davidov:2010:ESL:1944566.1944594}, the binary classification is a useful application and can be used as a filter that extracts sentiment sentences from a corpus for further processing. We also test how the performance is affected with and without using Bloom filters. The value $k$ for the $k$NN classifier is equal to 50. The results of the experiments are displayed in Table \ref{class_results}. In case of binary classification, the results depict the average score for all classes.

\begin{table}
\centering
\caption{Classification results for emoticons and hashtags (BF stands for Bloom filter and NBF for no Bloom filter)}
\label{class_results}
\resizebox{\linewidth}{!}{\begin{tabular}{|l|c|c|c|} \hline
 Setup & BF & NBF & Random baseline \\ \hline
 Multi-class Hashtags & 0.32 & 0.33 & 0.08\\ \hline
 Multi-class Emoticons & 0.55 & 0.56  & 0.25\\ \hline
 Binary Hashtags & 0.74 & 0.53  & 0.5\\ \hline
 Binary Emoticons & 0.77 & 0.69 &  0.5\\ \hline
\end{tabular}}
\end{table}

For multi-class classification the results are not very good but still they are way above the random baseline. We also observe that the results with and without the Bloom filters are almost the same. Thus, we deduce that for multi-class classification the Bloom filters marginally affect the classification performance. Furthermore, the outcome for emoticons is significantly better than hashtags which is expected due to the lower number of sentiment types. This behavior can also be explained by the ambiguity of hashtags and some overlap of sentiments. In case of binary classification there is a notable difference between the results with and without Bloom filters. These results may be somewhat unexpected but can be explicated when we take a look in Table \ref{no_match_fraction}. Table \ref{no_match_fraction} presents the fraction of test set tweets that are classified as neutral because of the Bloom filters and/or the weight threshold $w$ (no matching vectors are found). Notice that the integration of Bloom filters, leads to a bigger number of tweets with no matching vectors. Obviously, the excluded tweets have an immediate effect to the performance of the $k$NN classifier in case of binary classification. This happens since the number of tweets in the cross fold validation process is noticeably smaller compared to the multi-class classification. Overall, the results for binary classification with Bloom filters confirm the usefulness of our approach.

\begin{table}
\centering
\caption{Fraction of tweets with no matching vectors}
\label{no_match_fraction}
\begin{tabular}{|l|c|c|c|} \hline
 Setup & BF & NBF \\ \hline
 Multi-class Hashtags & 0.05 & 0.01\\ \hline
 Multi-class Emoticons & 0.05 & 0.02 \\ \hline
 Binary Hashtags & 0.05 & 0.03 \\ \hline
 Binary Emoticons & 0.08 & 0.06 \\ \hline
\end{tabular}
\end{table}

\subsection{Effect of k}

In this subsection, we attempt to alleviate the problem of our approach's low performance for binary classification without Bloom filters. To achieve this we measure the effect of $k$ in the classification performance of the algorithm. We test four different configurations where $k \in \{50, 100, 150, 200\}$. The outcome of this experimental evaluation is demonstrated in Table \ref{effect_of_k}. For both binary and multi-class classification, increasing $k$ affects slightly (or not at all) the harmonic f-score when we embody Bloom filters. The same thing does not apply when we do not use Bloom filters. More specifically, there is a great enhancement in the binary classification performance for hashtags and emoticons and a smaller improvement in case of multi-class classification. The inference of this experiment, is that larger values of $k$ can provide a great impulse in the performance of the algorithm when not using Bloom filters.

\begin{table}
\centering
\caption{Effect of $k$ in classification performance}
\label{effect_of_k}
\resizebox{\linewidth}{!}{\begin{tabular}{|l|c|c|c|c|} \hline
 Setup & $k=50$ & $k=100$ & $k=150$ & $k=200$ \\ \hline
 Multi-class Hashtags BF & 0.32 & 0.32 & 0.32 & 0.32 \\ \hline
 Multi-class Hashtags NBF & 0.33 & 0.35 & 0.37 & 0.37 \\ \hline
 Multi-class Emoticons BF & 0.55 & 0.55  & 0.55 & 0.55 \\ \hline
 Multi-class Emoticons NBF & 0.56 & 0.58  & 0.6 & 0.6 \\ \hline
 Binary Hashtags BF & 0.74 & 0.75  & 0.75 & 0.75 \\ \hline
 Binary Hashtags NBF & 0.53 & 0.62  & 0.68 & 0.72\\ \hline
 Binary Emoticons BF & 0.77 & 0.77 & 0.78 & 0.78\\ \hline
 Binary Emoticons NBF & 0.69 & 0.75 &  0.78 & 0.79\\ \hline
\end{tabular}}
\end{table}

\subsection{Space Compression}

\begin{figure}
\centering
\epsfig{file=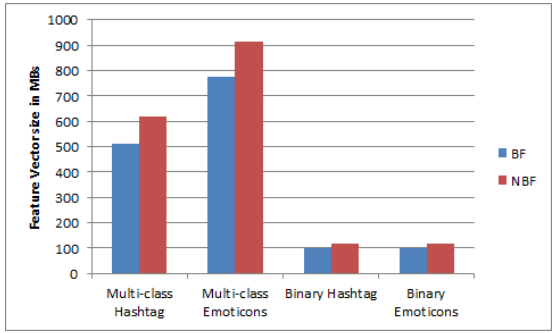,scale=.57}
\caption{Space compression of feature vector}
\label{compact_ratio}
\end{figure}

As stated and above, the Bloom filters can compact the space needed to store a set of elements, since more than one object can be stored to the bit vector. In this subsection, we elaborate on this aspect and present the compression ratio in the feature vectors when exploiting Bloom filters (in the way presented in Section 3.2) in our framework. The outcome of this measurement is depicted in Fig. \ref{compact_ratio}. In all cases, the Bloom filters manage to diminish the storage space required for the feature vectors by a fraction between 15-20\%. According to the analysis made so far, the importance of Bloom filters in our solution is twofold. They manage to both preserve a good classification performance, despite any errors they impose, and compact the storage space of the feature vectors. Consequently, we deduce that Bloom filters are very beneficial when dealing with large scale sentiment analysis data, that generate an exceeding amount of features during the feature vector construction step.

\subsection{Running Time and Scalability}

\begin{figure}
\centering
\epsfig{file=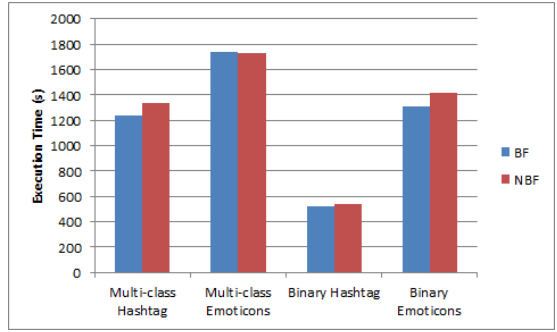,scale=.57}
\caption{Running time}
\label{running_time}
\end{figure}

In this final experiment, we compare the running time for multi-class and binary classification and measure the scalability of our approach. Initially, we calculate the execution time in all cases in order to detect if the Bloom filters speedup or slow down the running performance of our algorithm. The results when $k=50$ are presented in Fig. \ref{running_time}. It is worth noted that in the majority of cases, Bloom filters slightly boost the execution time performance. Despite needing more preprocessing time to produce the features with Bloom filters, in the end they pay off since the feature vector is smaller in size. This leads to lower I/O cost between the Map and Reduce tasks and consequently to less processing time. Multi-class classification for emoticons constitutes the only exception in our example.

\begin{figure}
\centering
\epsfig{file=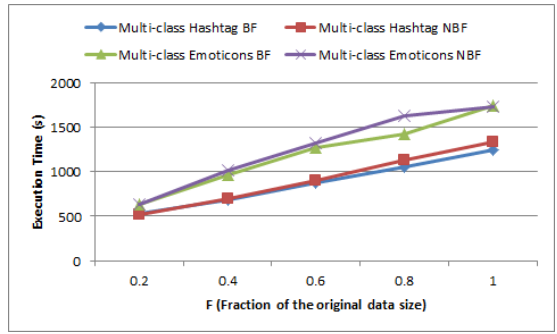,scale=.57}
\caption{Scalability}
\label{scalability}
\end{figure}

Finally, we investigate the scalability of our approach. We test the scalability only for the multi-class classification case since the produced feature vector in much bigger compared to the binary classification case. We create new chunks smaller in size that are a fraction $F$ of the original datasets, where $F \in $ \{0.2, 0.4, 0.6, 0.8\}. Moreover, we set the value of $k$ to 50. Figure \ref{scalability} presents the scalability results of our approach. From the outcome, we deduce that our algorithm scales almost linear as the data size increases in all cases. This proves that our solution is efficient, robust, scalable and therefore appropriate for big data sentiment analysis.

\section{Conclusions and Future Work}

In the context of this work, we presented a novel method for sentiment learning in the MapReduce framework. Our algorithm exploits the hashtags and emoticons inside a tweet, as sentiment labels, and proceeds to a classification procedure of diverse sentiment types in a parallel and distributed manner. Moreover, we utilize Bloom filters to compact the storage size of intermediate data and boost the performance of our algorithm. We conduct a variety of experiments to test the efficiency of our method. Through this extensive experimental evaluation we prove that our system is efficient, robust and scalable.

In the near future, we plan to extend and improve our framework by exploring more features that may be added in the feature vector and will increase the classification performance. Furthermore, we wish to explore more strategies for $F_H$ and $F_C$ bounds in order to achieve better separation between the HFWs and CWs. Finally, we plan to implement our solution in other platforms (e.g. Spark) and compare the performance with the current implementation.

\balance

\bibliographystyle{abbrv}
\bibliography{mrsat}
\end{document}